\documentclass[letterpaper]{mn2e}

\usepackage{rotating}
\usepackage{psfig}

\def\lsim{~\rlap{$<$}{\lower 1.0ex\hbox{$\sim$}}}
\def\bsim{~\rlap{$>$}{\lower 1.0ex\hbox{$\sim$}}}
\def\hmpc{\ {\rm {\it h}^{-1}Mpc}}

\def\hmmpc{\ {\rm {\it h}Mpc^{-1}}}
\def\hhhmmpc{\ {\rm {\it h}^{3}Mpc^{-3}}}

\def\mdh{\ {\rm M_\odot/{\it h}}}

\def\la{\langle}
\def\ra{\rangle}
\def\dd{{\rm d}}
\def\dc{\delta_{\rm c}}

\def\mathbi#1{\textbf{\em #1}}

\def\vx{\mathbi{x}}
\def\vk{\mathbi{k}}
\def\vq{\mathbi{q}}

\def\fnl{f_{\rm NL}}
\def\dh{\delta_{\rm h}}
\def\dm{\delta_{\rm m}}
\def\pmm{P_{\rm mm}}
\def\pmh{P_{\rm mh}}
\def\phh{P_{\rm hh}}
\def\bmh{b_{\rm mh}}
\def\bhh{b_{\rm hh}}
\def\pl{P_{_{\rm L}}}
\def\nh{\bar{n}_{\rm h}}

\def\etal{{\it et al.\ }}

\begin{document}

\title[non-Gaussian bias: theory vs. simulations]
      {Scale-dependent bias induced by local non-Gaussianity: 
       A comparison to N-body simulations}

\author[Desjacques \etal]
{Vincent Desjacques$^1$, Uro\v s  Seljak$^{1,2}$ and Ilian T. Iliev$^1$\\ 
$^1$ Institute for Theoretical Physics, University of Z\"urich, 
Winterthurerstrasse 190, CH-8057 Z\"urich, Switzerland \\ 
$^2$ Physics and Astronomy Department, University of California, and 
Lawrence Berkeley National Laboratory, \\ Berkeley, California 94720, USA }
\date{}
\pagerange{\pageref{firstpage}--\pageref{lastpage}}
\maketitle

\label{firstpage}

\begin{abstract}

We investigate the effect of primordial non-Gaussianity of the local
$\fnl$ type on the auto- and cross-power spectrum of dark matter
haloes using simulations of the $\Lambda$CDM cosmology.  We perform a
series of large N-body simulations of both positive and  negative
$\fnl$, spanning the range between 10 and 100.  Theoretical models
predict a scale-dependent bias correction $\Delta b(k,\fnl)$ that
depends on the linear halo bias $b(M)$. We measure the power spectra
for a range of halo mass and redshifts covering the relevant range of
existing galaxy and quasar populations. We show that auto and
cross-correlation analyses of bias are consistent with each other.  We
find that for low wavenumbers with $k < 0.03\hmmpc$ the theory and the
simulations  agree well with each other for biased haloes with
$b(M)>1.5$.  We show that   a scale-independent bias correction
improves the comparison between theory and  simulations on smaller
scales, where the scale-dependent effect rapidly becomes negligible.
The current limits  on $\fnl$ from Slosar et al. (2008) come mostly
from  very large scales $k <0.01\hmmpc$ and, therefore,  remain valid.
For the halo samples with $b(M)<1.5-2$ we find that the scale\--
dependent bias from non-Gaussianity  actually exceeds the theoretical
predictions.  Our results are consistent with the bias correction
scaling linearly with   $\fnl$.

\end{abstract}

\begin{keywords}
cosmology: theory --- gravitation --- dark matter ---  galaxies:
haloes ---
\end{keywords}

\section{Introduction}

Generic inflationary models based on the slow roll of a scalar field
predict a nearly scale-invariant and Gaussian spectrum of primordial  
curvature fluctuations (see Bartolo \etal 2004 for a review). 
While the latest measurements of the cosmic 
microwave background (CMB) anisotropies favour a slightly red power
spectrum (Komatsu \etal 2008), no significant detection of primordial
non-Gaussianity has been reported as yet from CMB and 
large-scale structures measurements. Nevertheless, improving 
the current limits would still strongly constrain mechanisms for the
generation of cosmological perturbations. 

Non-Gaussianity can be generated by nonlinearities in the relation
between the primordial curvature perturbation and the inflaton field
(e.g. Salopek \& Bond 1990; Gangui \etal 1994), interaction of scalar 
fields (e.g. Falk \etal 1993) or deviation from the (Bunch-Davies) 
ground state (e.g. Lesgourgues \etal 1997). 
A wide class of inflationary scenarios lead to non-Gaussianity of the
local type, which depends on the  local value of the potential
only. In these models, deviation from  Gaussianity can be conveniently
parametrised by a nonlinear coupling  parameter $\fnl$ through the
relation (e.g. Komatsu \& Spergel 2001) 
\begin{equation}
\Phi(\vx)=\phi(\vx)+\fnl\left(\phi(\vx)^2-\la\phi(\vx)^2\ra\right)\;,
\end{equation} 
where $\phi(\vx)$ is the Gaussian part of the curvature perturbation
in   the matter area. While single inflaton scenarios predict $\fnl$
much  less than unity, multi-field inflation models can generate
$\fnl\gg 1$ (Linde \& Mukhanov 1997; Lyth \etal 2003; Creminelli 2003;
Dvali \etal 2004; Zaldarriaga 2004; Arkani-Hamed \etal 2004;
Alishahiha \etal 2004). Alternatives to inflation, such as
cyclic/ekpyrotic model also predict large non-Gaussianity of  local
type (Creminelli \& Senatore 2007, Buchbinder \etal 2008, Lehners \&
Steinhardt 2008).

Higher order statistics of the curvature perturbation such as the
bispectrum can be computed straightforwardly from a perturbative
expansion of the homogeneous Robertson-Walker background
(e.g. Acquaviva \etal 2003; Maldacena 2003).  These statistics are
related to those of the CMB temperature anisotropy through the
radiation transfer function, which can be computed accurately using,
e.g., {\small CMBFAST} (Seljak \& Zaldarriaga 1996). Thus far,
analysis of the CMB bispectrum indicates that the data are fully
consistent with Gaussianity, with $|\fnl|\lsim 100$ (Komatsu \etal
2003; Creminelli \etal 2007; Komatsu \etal 2008; Smith \etal 2009;
see, however, Yadav \& Wandelt 2008 who report a detection at the
2.5$\sigma$ level), providing strong evidence for the quantum origin
of the primordial fluctuations.

Large-scale structures offer another route to test for the presence of
primordial non-Gaussianity. It has long been recognised that departure
from Gaussianity can significantly affect the high mass tail of the
dark matter halo distribution (Lucchin \& Matarrese 1988;
Colafrancesco \etal 1989; Chiu \etal 1998; Robinson \& Baker  2000;
Matarrese \etal 2000; Mathis \etal 2004, Kang \etal 2007;  Grossi
\etal 2007). Following this approach, X-ray cluster counts have been
used to constrain the amount of non-Gaussianity (e.g. Koyama \etal
1999, Robinson \etal 2000;  Willick 2000; Amara \& Refregier
2004). Galaxy clustering is also  sensitive to the statistical
properties of the primeval fluctuations.  Indeed, Grinstein \& Wise
(1986) pointed out early that primordial  non-Gaussianity could
significantly increase the amplitude of the  two-point correlation of
galaxies and clusters on large scales.  However, recent work has
mostly focused on higher order statistics such as  the bispectrum
(Scoccimarro \etal 2004; Sefusatti \& Komatsu 2007).

Dalal \etal (2008) have recently sparked renewed interest in the
clustering of rare objects by demonstrating the strong scale-dependent
bias arising from primordial non-Gaussianity of the local type.
It can be shown that the latter contributes a scale-dependent bias 
of the form (Dalal \etal 2008; Matarrese \& Verde 2008; Slosar \etal 
2008) 
\begin{equation}
\Delta b_\kappa(k,\fnl)=
3\fnl \left[b(M)-1\right]\dc\frac{\Omega_{\rm m}H_0^2}{k^2 T(k) D(z)}\;,
\label{eq:bshift}
\end{equation}
where $b(M)$ is the linear bias parameter, $H_0$ is the
Hubble  parameter, $T(k)$ is the matter transfer function, $D(z)$ is
the growth  factor normalised to $(1+z)^{-1}$ in the matter era and
$\dc \sim 1.68$ is the present-day (linear) critical density threshold.
While the  derivation
of this non-Gaussian bias correction presented in Dalal \etal (2008)
and Matarrese \& Verde (2008)   is strictly valid only for the highest
peaks of the density field, the peak-background split argument invoked
by Slosar \etal (2008) suggests that eq.~(\ref{eq:bshift}) should
apply to all peaks unrestrictedly, but is only valid in the limit of
long wavelength modes so that the background can be approximated as a
constant density.  Further work has confirmed the basic picture
(Afshordi \& Tolley 2008; McDonald 2008; Taruya \etal 2008).

Slosar \etal (2008) have applied eq.~(\ref{eq:bshift}) to constrain
the value of $\fnl$ using a compilation of large-scale structure data.
They find $-29<\fnl<+69$ (at 95\% confidence level). These limits are
competitive with those from WMAP5, $-9<\fnl<+111$ (Komatsu \etal 2008)
and $-4<\fnl< 80$ (Smith \etal 2009), demonstrating the promise of the
method. Future all sky surveys could achieve  constraints of the order
of $\fnl\sim 5-10$ (Dalal \etal 2008;  Afshordi \& Tolley 2008;
McDonald 2008; Carbone \etal 2008), assuming one knows how to extract
maximum information from the data (see, e.g., Slosar 2008).  In fact,
with sufficient high density of tracers it should be possible  to
circumvent the sampling variance (which is a serious issue since the
non-Gaussian effect is strongest on the largest scales) and alleviate
degeneracies with other  cosmological parameters, thereby allowing
for a potentially huge gains (Seljak 2008).

Still, in order to fully exploit the potential of forthcoming
large-scale  surveys, the method needs to be tested with large
numerical simulations.  Thus far, eq.~(\ref{eq:bshift}) has been
validated  only using the halo-matter  cross-power spectrum (Dalal
\etal 2008) and only  on very large scales, so its accuracy remains
uncertain. It is important to measure the effect in  the
auto-correlation of dark matter haloes, since the latter gives the
strongest constraint on $\fnl$ (Slosar \etal 2008). It is also
important to extend the analysis to smaller scales,  where the
peak-background split breaks down, as well as to less biased haloes.
The purpose of this paper is to address these issues in more
detail. We begin with a brief description of the N-body simulations
against which we calibrate the theory (\S\ref{sec:nbody}). Next, we
discuss the mass function and bias of the corresponding halo catalogues
and demonstrate the importance of including a scale-independent bias
correction in the comparison with the simulations
(\S\ref{sec:haloes}). The main body of the paper is \S\ref{sec:ngbias},
where we study in detail the impact of local non-Gaussianity on the
halo-matter and halo-halo power spectrum.  We conclude with a
discussion of the results in \S\ref{sec:conclusion}.

\section{The N-body simulations}
\label{sec:nbody}

Investigating the scale-dependence of the halo bias requires
simulations large enough so that many long wavelength modes are
sampled. At the same time, the simulations should resolve dark matter
haloes hosting luminous red  galaxies (LRGs) or quasars (QSOs), so
that one can construct halo samples whose statistical properties mimic
as closely  as possible those of the real data.

In this work, we use a series  of large N-body simulations of the
$\Lambda$CDM cosmology seeded with  Gaussian and non-Gaussian initial
conditions. The non-Gaussianity  is of the ``local'' form,
$\Phi=\phi+\fnl (\phi^2-\la\phi^2\ra)$, where $\Phi(\vx)$  is the
Bardeen potential.  It is important to note that this local
transformation is performed before multiplication by the matter
transfer function. $T(k)$ is computed with {\small CMBFAST} (Seljak \&
Zaldarriaga 1996) for the WMAP5 best-fitting parameters (Komatsu \etal
2008)~: $h=0.7$, $\Omega_{\rm m}=0.279$, $\Omega_{\rm b}=0.0462$,
$n_s=0.96$ and a normalisation of the curvature perturbations
$\Delta^2_{\cal R}=2.21\times 10^{-9}$ (at $k=0.02$Mpc$^{-1}$)  which
gives $\sigma_8\approx 0.81$. Five sets of three 1024$^3$ simulations,
each of which has $\fnl=0,\pm 100$, were run with the N-body code
{\small GADGET2} (Springel 2005).  We used the same Gaussian random
seed field $\phi$ in each set of runs so as to minimise the sampling
variance. We also explored lower values of $\fnl$ and ran 2 
realisations  
for each of the non-Gaussian models characterized by $\fnl=\pm 30$ and 
$\pm 10$. In all cases  
the box size is 1600$\hmpc$ with a force resolution of 0.04
times the mean interparticle distance.  The particle mass of these
simulations thus is $3.0\times 10^{11}\mdh$, enough to resolve haloes
down to $10^{13}\mdh$.

Haloes were identified using the {\small MPI} parallelised version of
the {\small AHF} halo finder which is based on the spherical
overdensity (SO) finder developed by Gill, Knebe \& Gibson
(2004). {\small AHF} estimates the local density around each halo
centre using a top-hat aperture. The virial mass  $M$ is defined by the
radius at which the inner overdensity exceeds $\Delta_{\rm vir}(z)$
times the background density $\bar{\rho}(z)$.  Note that $\Delta_{\rm
vir}(z)$ is an increasing  function of redshift ($\Delta_{\rm
vir}\approx 340$ at $z=0$).  We discard poorly resolved haloes and
only study those  containing at least 34 particles to reduce the error
in the mass  estimate (Warren \etal 2006). This implies a lower mass
limit $M=10^{13}\mdh$ which is about the typical mass of
QSO-hosting haloes at $1<z<2$ (e.g. Porciani \& Norberg 2006) and  is
a few times smaller than the mass of haloes harbouring LRGs in SDSS
(Mandelbaum etal. 2006). 

\begin{figure}
\center \resizebox{0.45\textwidth}{!}{\includegraphics{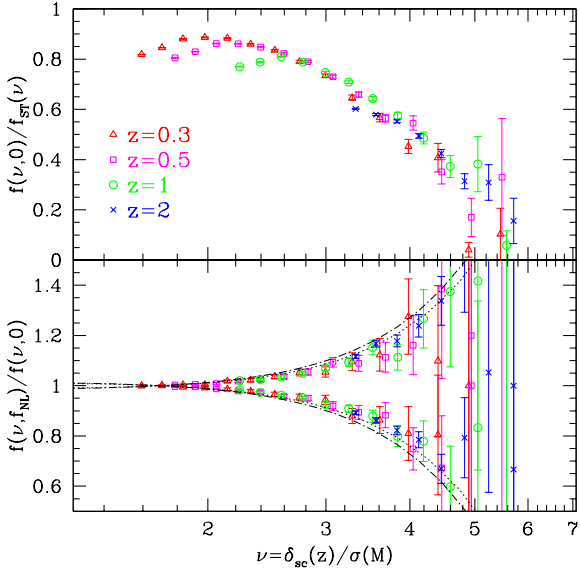}}
\caption{{\it Top panel}~: multiplicity function $f(\nu,0)$ for the
Gaussian simulations. Different symbols refer to different redshifts
as indicated. Results are shown relative to the Sheth-Tormen fitting
formula to emphasise deviation from the latter.  {\it Bottom panel}~:
ratio between the non-Gaussian and the  fiducial Gaussian mass
functions. The dotted and dotted-dashed curves  are the theoretical
prediction at $z=0$ and 2, which is based on an Edgeworth expansion of
the dark matter probability distribution function (see text). In both
panels,  error bars denote Poisson errors. For illustration,
$M=10^{15}\mdh$ corresponds to $\nu=3.2$, 5.2, 7.7 at redshift $z=0$,
1 and 2, respectively. Similarly, $M=10^{14}\mdh$ and $10^{13}\mdh$
correspond to $\nu=1.9$, 3, 4.5 and 1.2, 1.9, 2.9 respectively.}
\label{fig:fig1}
\end{figure}

\section{halo mass function and bias}
\label{sec:haloes}

\subsection{Multiplicity function}

Analytic arguments based on the Press-Schechter theory (Press-Schechter
1974; Bond \etal 1991; Sheth \& Tormen 1999) predict that the halo
mass function $n(M,z)$ is entirely specified by the distribution 
$\nu f(\nu)$ of first-crossings, or multiplicity function 
\begin{equation} 
\nu f(\nu)=M^2\,\frac{n(M,z)}{\bar{\rho}}\frac{\dd\ln M}{\dd\ln\nu}\;.
\label{eq:fnu}
\end{equation} 
The peak height $\nu(M,z)=\dc(z)/\sigma(M)$, where $\dc(z)\approx
1.68 D(0)/D(z)$  is the critical linear overdensity for collapse (assumed
spherical throughout this paper), is the typical amplitude of
fluctuations that produce haloes of mass $M$ by redshift $z$.  A
characteristic mass for clustering, $M_\star(z)$, can then be defined
through $\nu(M,z)=1$. For the present cosmology, $M_\star(0)\approx
3.5\times 10^{12}\mdh$.

The top panel of Fig.~\ref{fig:fig1} shows the multiplicity function
of the SO haloes extracted from the Gaussian simulations at redshift
$z=0.3$, 0.5, 1 and 2.  The numerical data are plotted with respect
to the Sheth-Tormen function (Sheth-Tormen 1999) to emphasise the
large deviation from the latter.  This departure is, however, not
really surprising since the Sheth-Tormen  formula is a fit to the mass
function of Friends\--of\--Friends haloes  (extracted from the GIF
simulations, see Kauffmann \etal 1999). We have not attempted to fit
the multiplicity function of our SO haloes given the limited volume
and  dynamic range of our simulations. Instead, we have found more
useful to  assess whether the impact of local non-Gaussianity  on the
halo mass function is consistent with theoretical expectations.

To test this we have plotted the ratio $f(\nu,\fnl)/f(\nu,0)$ in the
bottom panel of Fig.~\ref{fig:fig1} for the simulations with $\fnl=\pm
100$.  The presence of primordial non-Gaussianity enhances or
suppresses the high peak tail of the multiplicity function depending
on the sign of $\fnl$. As recognised in previous papers
(e.g. Matarrese \etal 2000; Sefusatti \etal  2007), despite the lack
of a  reliable Gaussian mass function, deviations from Gaussianity can
be  modelled analytically using the Press-Schechter formalism.  Here
we follow the simple extension introduced by LoVerde \etal (2008; see
also Chiu \etal 1998) and replace the Gaussian probability
distribution function (PDF) of the density field by the generic
Edgeworth expansion (e.g., Scherrer \&  Bertschinger 1991; Juskiewicz
\etal 1995). Neglecting cumulants  other than the skewness
$S_3(M)=\la\delta^3_M\ra/\la\delta^2_M\ra^2$ and truncating the series
expansion at $S_3$, the non-Gaussian correction factor reads 
(LoVerde \etal 2008)
\begin{equation}
\frac{f(\nu,\fnl)}{f(\nu,0)}=
1+\frac{1}{6}\,\sigma S_3 \left(\nu^3-3\nu\right)-\frac{1}{6}\,
\frac{\dd(\sigma S_3)}{\dd\ln\nu}\left(\nu-\frac{1}{\nu}\right)
\label{eq:ngfnu}
\end{equation} 
after integration over regions above the critical density for
collapse. Note that we have omitted the explicit redshift dependence.
Strictly speaking however, the ratio $f(\nu,\fnl)/f(\nu,0)$ depends
distinctly upon the variables  $M$ (or $\nu$) {\it and} $z$ due to the
presence of $\sigma S_3(M)$. Our notation is motivated by the fact
that the measured non-Gaussian correction, as plotted in the bottom
panel of Fig.~\ref{fig:fig1}, appears to depend  mostly on the peak
height.

Equation~(\ref{eq:ngfnu}) requires knowledge of the skewness $S_3(M)$
of the smoothed density field $\delta_M$, which we compute analytically 
using the relation (see Appendix~\S\ref{app:PT})
\begin{eqnarray}
\sigma^4 S_3(M) \!\!\! &=& \!\!\! \frac{\fnl}{(2\pi^2)^2}\int_0^\infty
\!\!\dd k_1\,k_1^2 \alpha(M,k_1) P_\phi(k_1) \label{eq:skew} \\ 
&& \times \int_0^\infty\!\!\dd k_2\,k_2^2 \alpha(M,k_2) P_\phi(k_2) 
\nonumber \\ 
&& \times \int_{-1}^{+1}\!\!\dd\mu\,\alpha(M,k)\left[1+2\frac{P_\phi(k)}
{P_\phi(k_2)}\right] \nonumber \;,
\end{eqnarray}
where $k^2=k_1^2+k_2^2+2\mu k_1 k_2$, $P_\phi(k)$ is the power spectrum 
of linear curvature perturbations in the matter-dominated era,
\begin{equation}
\alpha(M,k)=\frac{2}{3\Omega_{\rm m}H_0^2}
D(z) k^2 T(k) W(M,k)
\label{eq:alpha}
\end{equation} and $W(M,k)$ is a (spherically symmetric) window
function of characteristic mass scale $M$. Over the mass range probed
by our simulations, $10^{13}\lsim M\lsim 5\times 10^{15}\mdh$, $\sigma
S_3(M)$ is a monotonic decreasing function of $M$ that varies in the
narrow range  $\sim 3-3.3\times 10^{-4}\fnl$ for the top-hat filter
assumed here.  Furthermore, the $\sigma S_3$ term dominates the total 
contribution to the non-Gaussian correction when the peak height is 
$\nu\bsim 2$.

The resulting non-Gaussian correction is plotted  in the bottom panel
of Fig.~\ref{fig:fig1} for two different redshifts, $z=0$ (dotted) and
2 (dotted-dashed). The truncated expansion  eq.~(\ref{eq:ngfnu})
agrees reasonably  well with the numerical data, suggesting thereby
that cumulants higher  than $S_3$ may not be important in the range of
mass and redshift  considered here. Note also that for positive
$\fnl$ the mass function is enhanced more at the high mass end and
that  this is similar to an increase in the amplitude of fluctuations
$\sigma_8$.  Hence, $\fnl$ is somewhat degenerate with  $\sigma_8$
since, in both cases, the effect increases with mass  (compare with
Fig.3 of Mandelbaum \& Seljak 2007 for instance).  However, at
$\nu=3.2$ (i.e. $M=10^{15}M_{\sun}/h$ at $z=0$) the increase in mass
function for $\fnl=100$ is 15\%, which corresponds  to less than 0.01
change in $\sigma_8$. Therefore, given the current uncertainties in
the cluster abundance (which translate into 0.03 error on $\sigma_8$,
Vikhlinin \etal 2008),  the prospects of  using halo mass function to
place competitive limits on $\fnl$ with the current data are small. 

\begin{figure}
\center \resizebox{0.45\textwidth}{!}{\includegraphics{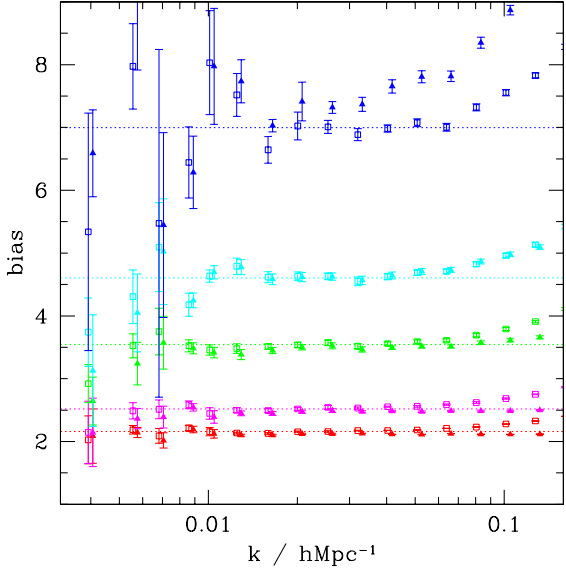}}
\caption{Halo bias as a function of wavenumber. Results are shown at
redshift $z=0.3$, 0.5, 1, 1.4 and 2 (from bottom to top) for haloes
with mass above  $2\times 10^{13}\mdh$. Filled and empty symbols
represent the bias estimators $\bhh=\sqrt{\phh/\pmm}$ and
$\bmh=\pmh/\pmm$, respectively. The bins are equally spaced  in
logarithmic space with a bin width $\Delta\log k=0.1$. Measurements of
$\bhh$ have been slightly shifted horizontally for clarity. The
horizontal lines indicate our  estimate of the linear bias $b(M)$ (see
text). Notice that $\phh(k)$ is corrected for shot-noise.}
\label{fig:fig2}
\end{figure}

\subsection{Linear bias}

Having checked that the level of non-Gaussianity in the mass  function
is consistent with simple theoretical expectations, we now turn to the
clustering  of dark matter haloes.

We interpolate the dark matter particles and halo centres onto a
regular cubical mesh. The resulting dark matter and halo fluctuation
fields, $\dm(\vk)$  and $\dh(\vk)$, are then Fourier transformed to
yield the matter-matter,  halo-matter and halo-halo power spectra
$\pmm(k)$, $\pmh(k)$ and  $\phh(k)$, respectively.  Notice that the
power spectra are computed  on a 512$^3$ grid to reduce the
computational expenses. Still, the Nyquist wavenumber is sufficiently
large, $\approx 1\hmmpc$, to allow for an accurate measurement of the
power in wavemodes of amplitude $k\lsim 0.1\hmmpc$. As we will see
shortly,  the impact of local non-Gaussianity is negligible at
$k=0.1\hmmpc$, but increases rapidly with decreasing wavenumber.

On linear scales, the halo bias $b(\vk)=\dh(\vk)/\dm(\vk)$ approaches
a constant, albeit mass-dependent value $b(M)$.  The linear halo bias
$b(M)$ needs to be measured accurately as it  controls the strength of
the scale-dependent bias correction induced by local  non-Gaussianity. 
To proceed, we may consider the following estimates of $b(k)$ for a 
given halo sample,
\begin{equation}
\bhh(k)=\sqrt{\frac{\phh(k)}{\pmm(k)}},~~~
\bmh(k)=\frac{\pmh(k)}{\pmm(k)}\;.
\end{equation}
In the following, we will always correct the halo power spectrum for
shot-noise, which we assume to be  $1/\nh$ if dark matter  haloes are
a Poisson sampling of some continuous  field.  While this discreteness
correction is negligible for $\pmm(k)$ and $\pmh(k)$ due to the large
number of dark matter particles, it can be quite significant for
$\phh(k)$.

\begin{figure}
\center \resizebox{0.45\textwidth}{!}{\includegraphics{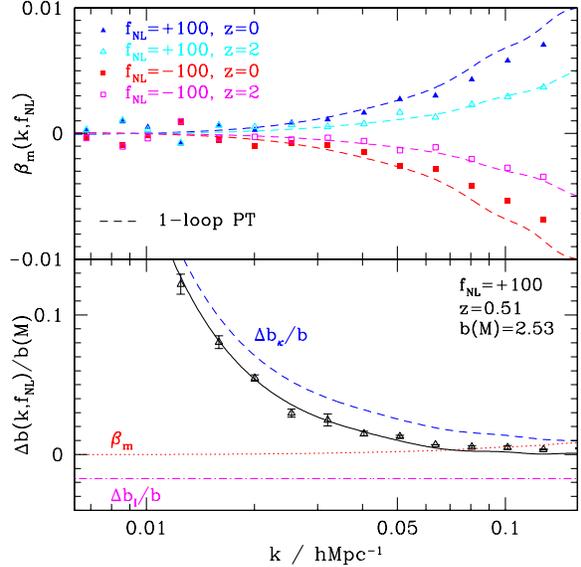}}
\caption{{\it Top panel}~: Non-Gaussian correction  $\beta_{\rm
m}(k,\fnl)=\Delta\pmm(k,\fnl)/\pmm(k,0)$ to the matter power spectrum
that originates from primordial non-Gaussianity of the local type.
Results are shown at redshift $z=0$ and 2 for $\fnl=\pm 100$. The
dashed  curves indicate the prediction from a leading-order
perturbative expansion.  {\it Bottom panel}~: Non-Gaussian bias
correction for the haloes of mass $M>2\times 10^{13}\mdh$ extracted
from the snapshot at  $z=0.5$ (filled symbols). The solid curve
represents our theoretical model eq.~(\ref{eq:dbias}). The  dashed,
dotted and dashed-dotted curves show the three separate  contributions
that arise at first order in $\fnl$. Our theoretical scaling agrees
very well with the data for $k\lsim 0.05\hmmpc$.}
\label{fig:fig3}
\end{figure}

In Fig.~\ref{fig:fig2}, the result of measuring $\bmh(k)$ and
$\bhh(k)$ in the Gaussian simulations is shown at various redshifts
for the haloes of mass   $M>2\times 10^{13}\mdh$. Error bars indicate
the scatter among the various realisations.  Except for the
most  biased sample, $\bmh(k)$ and $\bhh(k)$ are nearly constant and
agree well with each other when the wavenumber varies in the
``linear'' range  $\sim 0.005-0.05\hmmpc$. On these scales, the slight
offset between $\bmh(k)$ and $\bhh(k)$ suggests that the shot-noise
correction $1/\nh$ might be too large for the low bias haloes and too
small for the highest bias halo. It is worth pointing out that the
hypothesis  of shot noise being $1/\nh$ for the dark matter haloes
remains unproven  (McDonald 2008) and it is an issue worth exploring
further.   Here we will be mostly looking at ratios of power spectra
with and without  non-Gaussianity, so this is less of an issue.  Both
bias quantities feature some scale-dependence on smaller scales,
$k\bsim 0.05-0.1\hmmpc$. This is best seen in the most biased sample.
We will use $\bmh(k)$ as a proxy for the linear halo bias since it is
less sensitive to shot-noise. In Fig.~\ref{fig:fig2}, the horizontal
lines indicate our fit to $b(M)$ obtained from the measurement of
$\bmh(k)$ at wavenumber $0.005<k<0.05\hmmpc$.

\subsection{Non-gaussian bias shift}

As shown in Dalal \etal (2008), Matarrese \& Verde (2008) and Slosar
\etal (2008), local non-Gaussianity gives rise to the scale-dependent
bias correction eq.~(\ref{eq:bshift}). However,  at the lowest order there
are  two additional, albeit relatively smaller, corrections which
arise from the dependence of both the halo number density  $n(M,z)$
and the matter power spectrum $\pmm$ on $\fnl$. As we will see
shortly,  the inclusion of these extra terms substantially improves
the comparison between the theory and the simulations.

Firstly, assuming the peak-background split holds, the change in  the
mean number density of haloes induces a scale-independent shift which
we denote by $\Delta b_{\rm I}(\fnl)$.  The existence of such a term
was noted in Slosar \etal (2008) and  Afshordi \& Tolley (2008).
Using the non-Gaussian fractional correction eq.~(\ref{eq:ngfnu}) 
(which is not universal), this  contribution reads
\begin{eqnarray}
\lefteqn{\Delta b_{\rm I}(\fnl)
=-\frac{1}{\sigma}\frac{\partial}{\partial\nu}\ln\left(\frac{f(\nu,\fnl)}
{f(\nu,0)}\right)} \label{eq:bi} \\
&& = -\frac{f(\nu,0)}{6\sigma f(\nu,\fnl)}\,
\left[3\,\sigma S_3\left(\nu^2-1\right)-\frac{\dd^2(\sigma S_3)}{\dd\ln\nu^2}
\left(1-\frac{1}{\nu^2}\right)\right. \nonumber \\
&& ~~~~~\left.+\frac{\dd(\sigma S_3)}{\dd\ln\nu}\left(\nu^2-4-\frac{1}{\nu^2}
\right)\right] \nonumber\;.
\end{eqnarray}
This approximation should work reasonably well for moderate values
of the peak height, $\nu\lsim 4$, for which the formula of LoVerde
\etal (2008) matches well our data (see Fig.~\ref{fig:fig1}). It is
worth noticing that $\Delta b_{\rm I}(\fnl)$ has a sign opposite to
that of $\fnl$ (because the  bias decreases when the mass function
goes up). In practice, to estimate $\Delta b_{\rm I}(\fnl)$ for a
given halo sample, we evaluate $\sigma S_3$ and $\nu$ at the scale
corresponding to the average halo mass $\bar{M}$ of the sample.
Furthermore, since we consider only first order corrections to the
Gaussian bias, we set $f(\nu,0)=f(\nu,\fnl)$ in the above 
expression so that $\Delta b_{\rm I}$ is truly first order in $\fnl$.

\begin{figure}
\center \resizebox{0.45\textwidth}{!}{\includegraphics{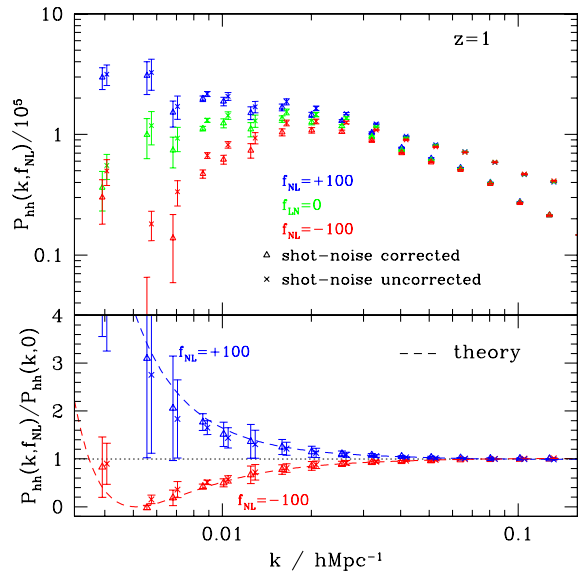}}
\caption{{\it Top panel}~: A comparison between the auto-power
spectrum with and without the shot-noise correction.  $P_{\rm
hh}(k,\fnl)$  is measured at $z=1$ for haloes of mass $M>2\times
10^{13}\mdh$. From top to bottom, the various symbols represent
the simulation results with $\fnl=+100$ (blue), 0 (green) and
-100 (red). The linear bias of this sample is $b(M)\approx
2.5$. {\it Bottom panel}~:  $P_{\rm hh}(k,\fnl)/P_{\rm hh}(k,0)$ as a
function of wavenumber.  The dashed curves denote the theoretical
prediction (see text).  In both panels, measurements without the
shot-noise correction have been shifted horizontally for clarity.}
\label{fig:fig4}
\end{figure}

Secondly, primordial non-Gaussianity affects the matter power spectrum
as  positive values of $\fnl$ tend to increase the small-scale power
(Scoccimarro \etal 2004; Grossi \etal 2008; Taruya \etal 2008).  For
$\fnl\sim {\cal O}(10^2)$, the magnitude of this correction is at a
per cent level in the weakly nonlinear regime $k\lsim 0.1\hmmpc$. In
order to illustrate this effect, the top panel of Fig.~\ref{fig:fig3}
displays the deviation $\beta_{\rm
m}(k,\fnl)=\Delta\pmm(k,\fnl)/\pmm(k,\fnl=0)$ that arises from the
presence of primordial non-Gaussianity of the local type. The symbols
show the result of measuring this ratio from the snapshots at redshift
$z=0$ and 2, whereas the dashed curves show the prediction from
one-loop perturbation  theory (Taruya \etal 2008; see also
Appendix~\S\ref{app:PT}). As we can see, leading order  perturbation
theory (PT) provides an excellent description of the effect over the
wavenumbers of interest, $k\lsim 0.1\hmmpc$. At $z=0$, one-loop PT
overestimates the non-Gaussian correction by $\sim 15$  per cent for
$k=0.1\hmpc$ and it is possible the agreement could be improved
further using renormalised perturbation theory (see, e.g., Crocce \&
Scoccimarro 2008). 

Summarizing, local non-Gaussianity adds a correction $\Delta b(k,\fnl)$ 
to the bias $b(k)$ of dark matter haloes that can be written as
\begin{equation}
\Delta b(k,\fnl)=\Delta b_\kappa(k,\fnl)+\Delta b_{\rm I}(\fnl)
+b(M)\beta_{\rm m}(k,\fnl)
\label{eq:dbias}
\end{equation}
at first order in $\fnl$.  The bottom panel of Fig.~\ref{fig:fig3}
illustrates the relative  contribution of these terms for haloes of
mass $M>2\times 10^{13}\mdh$ identified at redshift $z=0.5$. The solid
curve shows the total  non-Gaussian bias $\Delta b(k,\fnl)$.
Considering only the scale-dependent shift $\Delta b_\kappa$ leads to
an apparent suppression of the effect in simulations relative to the
theory.  Including the scale-independent  correction $\Delta b_{\rm
I}$  considerably improves the agreement at wavenumbers $k\lsim
0.05\hmmpc$.  Finally, adding the scale-dependent term
$b(M)\beta_{\rm m}$ further adjusts  the match at small scale $k\bsim
0.05\hmmpc$ by making the non-Gaussian  bias shift less negative.

\section{Results}
\label{sec:ngbias}

In order to quantify the effect of non-Gaussianity on the halo bias, 
we will consider the ratios~\footnote{Strictly speaking, 
$\phh(k,\fnl)/\phh(k,0)$ is equal to 
$[1+(\Delta_\kappa+\Delta_{\rm I})/b]^2+\beta_{\rm m}-1$,
which differs from eq.~(\ref{eq:ratio}) by $\beta_{\rm m}+{\cal O}
(\beta_{\rm m}^2)$. In what follows however, we will use 
$(1+\Delta b/b)^2-1$ for notational convenience.}
\begin{eqnarray}
\frac{\pmh(k,\fnl)}{\pmh(k,0)}-1 \!\!\! &=& \!\!\!
\frac{\Delta b(k,\fnl)} {b(M)} \label{eq:ratio} \\
\frac{\phh(k,\fnl)}{\phh(k,0)}-1 \!\!\! &=& \!\!\! 
\left(1+\frac{\Delta b(k,\fnl)} {b(M)}\right)^2-1 \nonumber\;.
\end{eqnarray}
Moreover, we shall also quantify the departure from the theory as a
function of wavemode amplitude with the ratio $\Delta b^{s}/\Delta
b^{t}$. Here,  $\Delta b^{s}$ is the non-Gaussian bias correction
measured from the simulation whereas $\Delta b^{t}$ is the theoretical
scaling eq.~(\ref{eq:dbias}).

\begin{figure*}
\center
\resizebox{0.9\textwidth}{!}{\includegraphics{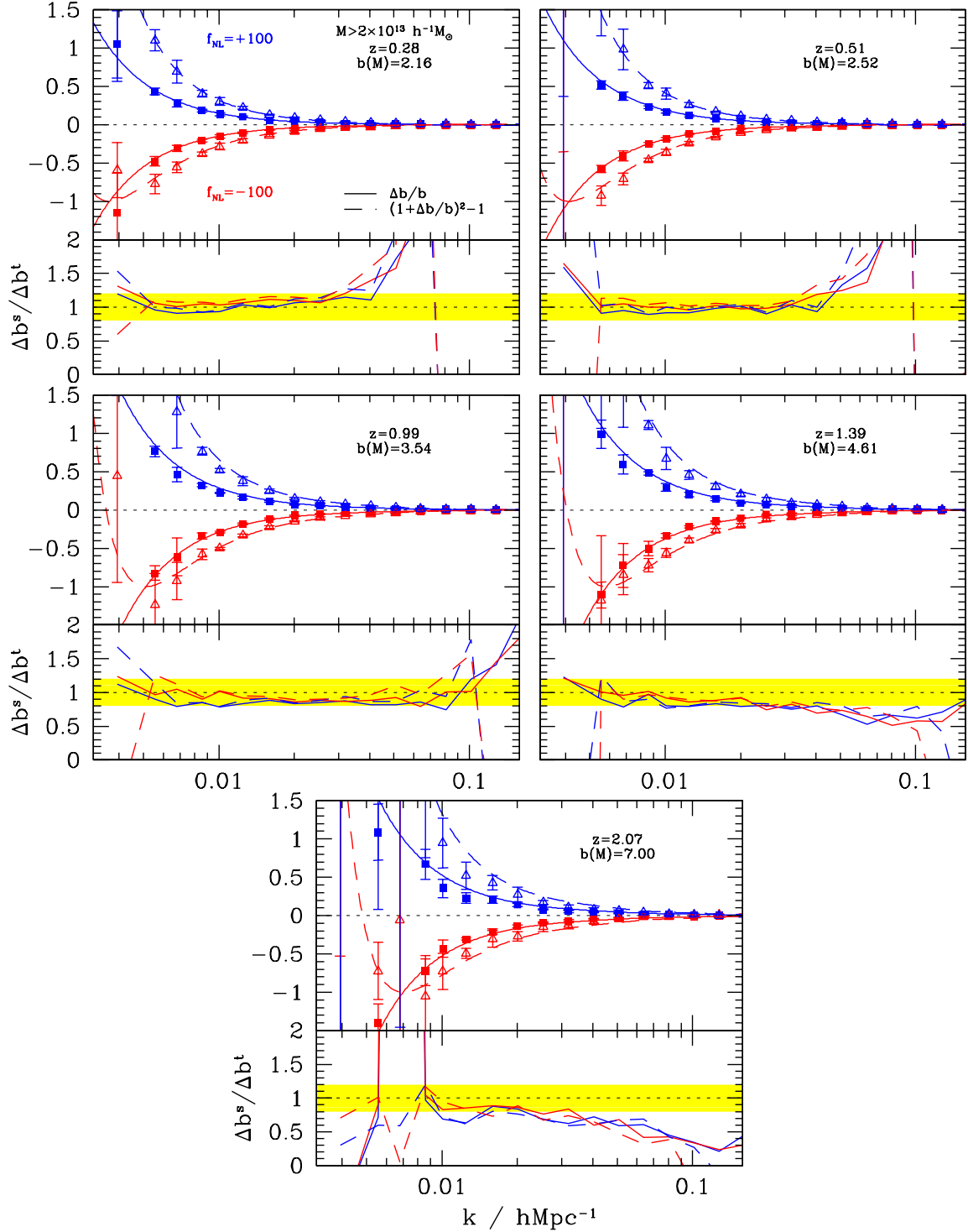}}
\caption{Non-Gaussian bias correction measured in the simulations at
various redshifts for haloes of mass $M>2\times 10^{13}\mdh$ (colors
as in Fig.~\ref{fig:fig4}). In each panel, the upper plot shows  the
ratio $P_{\rm hh}(k,\fnl)/P_{\rm hh}(k,0)-1$ (dotted curves, empty
symbols) and  $P_{\rm mh}(k,\fnl)/P_{\rm mh}(k,0)-1$ (solid curves,
filled symbols).   The error bars represent the scatter among 5
realisations. The respective output redshift and linear halo bias are
also quoted. The bottom of each panel displays the departure from the
theoretical prediction, $\Delta b^{s}/\Delta b^{t}$ (see text). The
shaded area indicates the domain where the deviation is  less than 20
percent. The theory agrees reasonably well with the measurements at 
wavenumber $k\lsim 0.03\hmmpc$.}
\label{fig:fig5}
\end{figure*}

\begin{figure*}
\center 
\resizebox{0.9\textwidth}{!}{\includegraphics{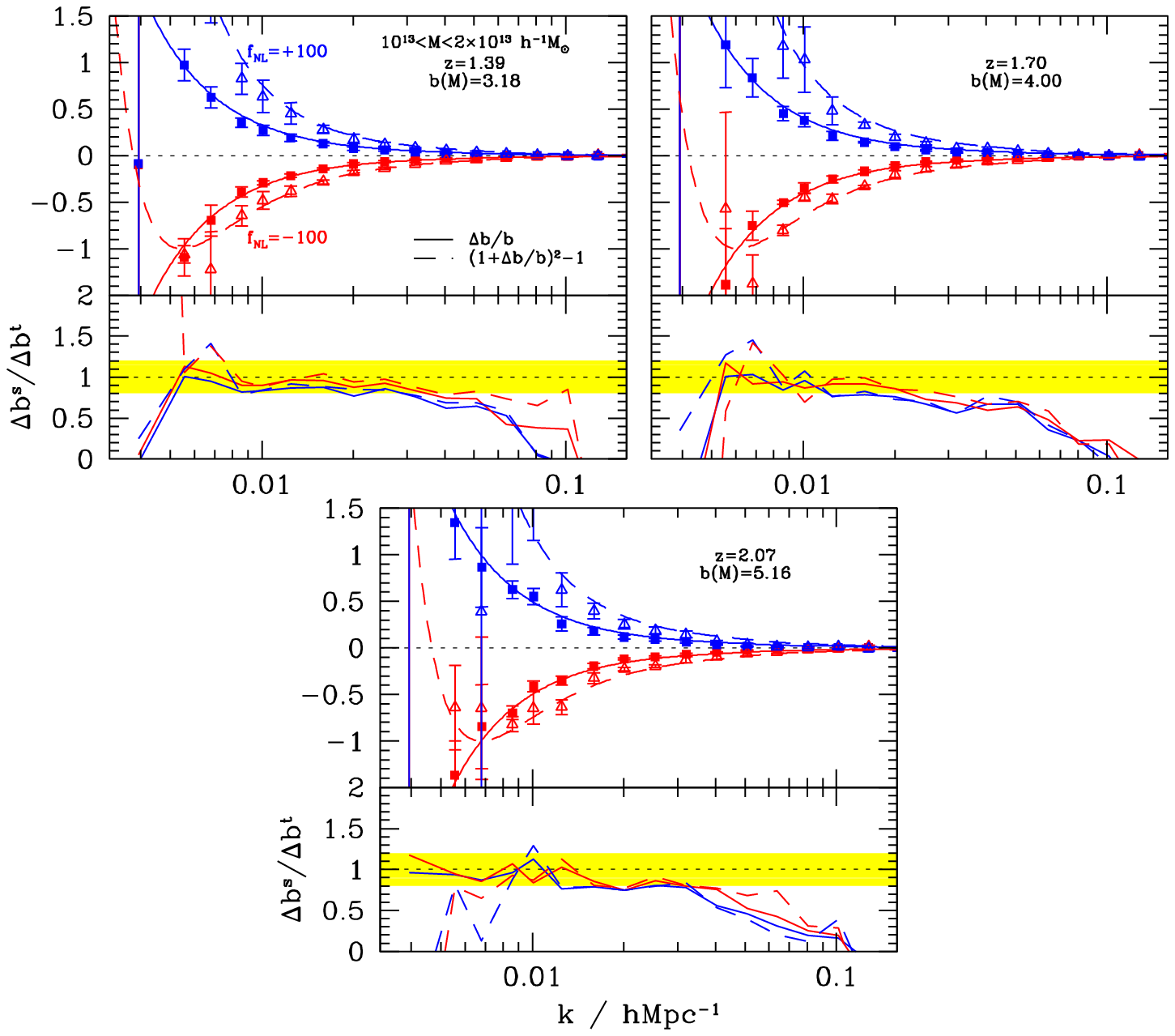}}
\caption{Same as Fig.~\ref{fig:fig5} but for haloes extracted from
the simulation outputs at $z=1.4$, 1.7 and 2, with a mass in the range
$1<M<2\times 10^{13}$. The halo sample at $z=1.4$ is close to 
the QSO sample used by Slosar \etal (2008), for which $z=1.8$ 
and $b=2.7$.}
\label{fig:fig6}
\end{figure*}

Before proceeding we look at the effect of the shot-noise correction
on the measurement of the non-Gaussian bias $\Delta b(k,\fnl)$.  In
Fig.~\ref{fig:fig4}, the averaged halo power spectrum and the ratio
$\phh(k,0)/\phh(k,\fnl)$ are shown before and after applying the
discreteness correction. Error bars represent the scatter among the
realisations. The bias and the number  density of the halo
sample considered here is $b(M)\approx 2.5$ and  $\nh\approx
10^{-4}\hhhmmpc$, respectively.

As we can see, the shot-noise can have a non-negligible effect on the
largest scales, specially for the haloes extracted from the
simulations with $\fnl=-100$ for which the large scale power crosses
zero on very  large scales.   For this particular sample, the
shot-noise correction enhances the measurement of $\Delta b(k,\fnl)$
by 10-15 per cent at scales $k\lsim 0.03\hmmpc$, regardless of the
sign of $\fnl$. While the exact amount of correction depends upon the
bias and the number density of the halo sample under consideration, it
is clear that any attempt to measure  $\Delta b(k,\fnl)$ at the few
per cent level must include the discreteness correction.

\subsection{Non-Gaussian bias from the halo-halo and halo-matter
power spectra}

We have measured power spectra for a range of halo masses and
redshifts, covering the relevant range of statistical properties
corresponding to the available data sets of galaxies or quasar
populations with different luminosities and bias.  The results are
summarised in Figures~\ref{fig:fig5} and \ref{fig:fig6}, where the
averaged $\Delta b/b$ and $(1+\Delta b/b)^2-1$ are plotted as a
function of wavenumber.  The deviation from the theoretical
prediction, $\Delta b^{s}/\Delta b^{t}$, is also shown at the bottom
of each panel. The shaded region indicates a deviation less than 20
per cent. To reduce the impact of sampling variance, we first compute
the ratios $\pmh(k,\fnl)/\pmh(k,0)$ and $\phh(k,\fnl)/\phh(k,0)$ for
each realisation, and then average over the realisations (see, e.g.,
Smith \etal 2007). We note that reversing the sequence of operations,
i.e. taking the ratio of averaged power spectra, gives very similar
average values.  Error bars denote the scatter around the mean and,
therefore, may underestimate the true errors since they are computed
from a small number of realisations.

\begin{figure*}
\center 
\resizebox{0.9\textwidth}{!}{\includegraphics{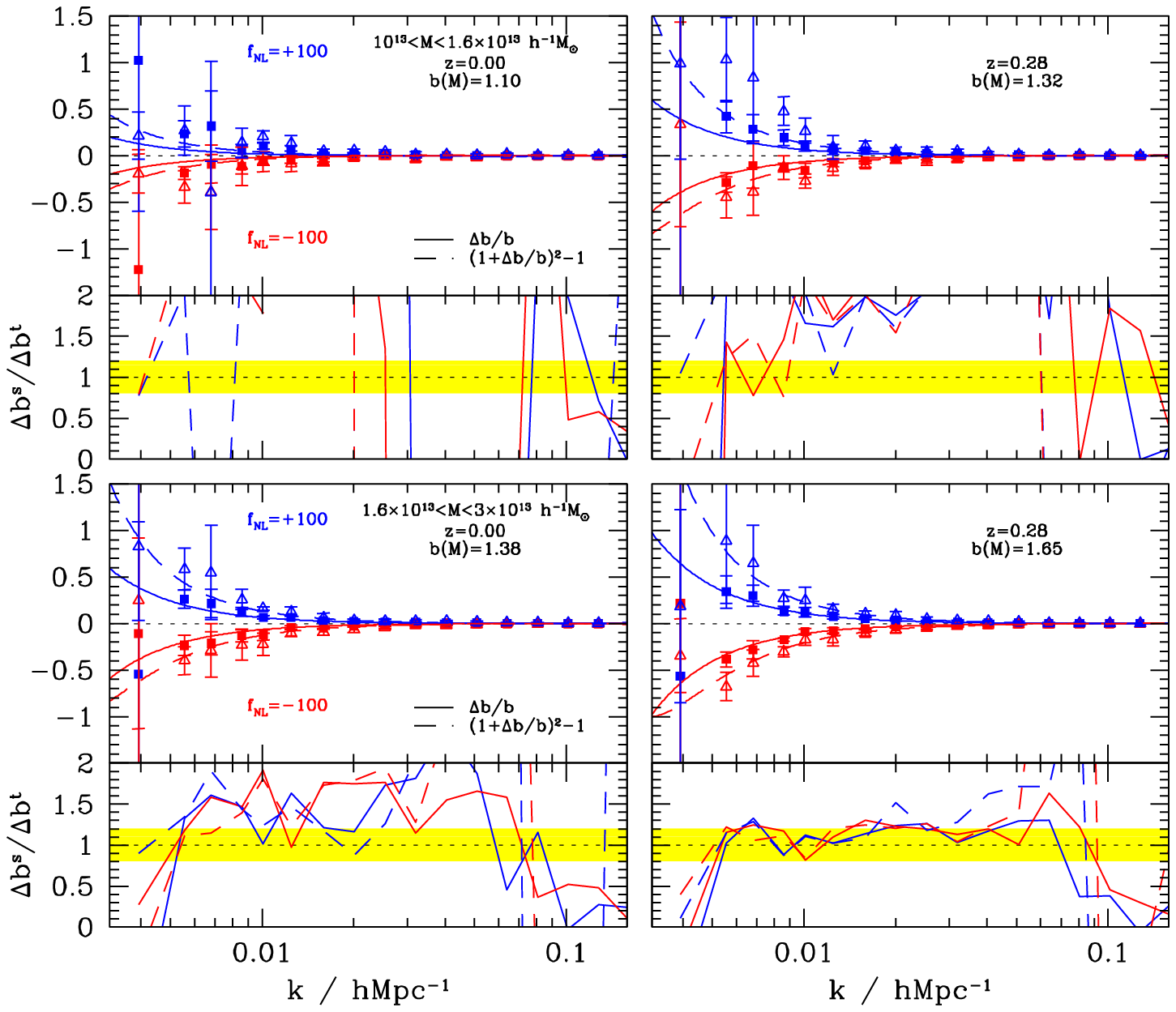}}
\caption{Same as Fig.~\ref{fig:fig5} but for haloes extracted from the
simulation outputs at $z=0$ and 0.3, with a mass in the range
$1<M<1.6\times 10^{13}$ (upper panels) and  $1.6<M<3\times
10^{13}\mdh$ (lower panels). The sample with $z=0.28$ and $b=1.65$
roughly corresponds to  LRG sample used by Slosar \etal (2008).}
\label{fig:fig7}
\end{figure*}

As we can see, the theoretical prediction provides a very good
description of the simulations at small wavenumber $k\lsim
0.03\hmmpc$, but the ratio $\Delta b^{s}/\Delta b^{t}$  differs
significantly from unity at larger wavenumbers.  The exact amount of
deviation depends weakly on the sign of $\fnl$. For moderately biased
haloes with $2<b(M)< 3$ the theory approaches the numerical  results
already on scale $k\lsim 0.05\hmmpc$. For the highly biased samples
$b>3$,  the theory overpredicts the effect seen in simulations on all
scales, but somewhat more on smaller scales, although in the high bias
limit   the numerical data is noisier due to the very low number
density of haloes.  It is worth noticing that, at the largest scales
$k\lsim 0.005\hmmpc$, the cross-power spectrum $\pmh(k,\fnl=-100)$
goes negative while $\phh(k,\fnl=-100)$ remains positive and even
increases,  in good agreement with the analytic prediction.  

We suspect these deviations at high wavenumber are mostly due to the
breakdown of the peak-background split  approximation which was used
in the derivation of the scale-dependent bias term $\Delta b_\kappa$
in Slosar \etal (2008). For this approximation to be valid one assumes
that  the long wavelength modes act as a homogeneous change of the
background,  from which the effect of the non-Gaussianity is computed
by comparing it to the local rescaling of the fluctuation amplitude.
Clearly this assumption breaks down once the wavelength of the mode
becomes  small. Uncertainties in the scale-independent correction also
affect $\Delta b^s/\Delta b^t$.  In this paper we use analytic
predictions based on equation (\ref{eq:bi}),  but we could also treat
the scale-independent bias as a free parameter that we fit to the
data, as done in the actual data analysis of  Slosar \etal 2008. For
example,  a $\sim 20$ per cent smaller (larger) $\Delta b_{\rm I}$ at
$b(M)\lsim 3$ ($b(M)\bsim 3$) would noticeably improve the convergence
at large $k$.  Finally, notice that the auto- and cross-power spectrum
of haloes give comparable results at all but the (poorly sampled)
largest scales, where  sampling variance prevents us from making any
conclusions.  This confirms the validity of the analysis in Slosar
\etal (2008), where  this effect was applied to the auto-correlations
of galaxies and quasars.

To assess the extent to which the agreement between simulation and
theory depends upon the halo mass and bias,  Figures~\ref{fig:fig6}
and ~\ref{fig:fig7} further explore the effect in the low and high
redshift outputs.  In Fig.~\ref{fig:fig6}, the non-Gaussian bias is
shown for haloes that correspond more closely to the quasars used by
Slosar \etal (2008), which are at $z=1.8$ and with $b=2.7$. Our halo
samples span a similar redshift range, $1.4<z<2$. However, the mass
cut $10^{13}<M<2\times 10^{13}\mdh$ gives larger values of the bias,
$3\lsim b(M)\lsim 5$,  suggesting that the quasars are hosted by
haloes (slightly) less massive than $10^{13}\mdh$  (unresolved in our
simulations).  As can be seen, the correction factor $\Delta
b^{s}/\Delta b^{t}$ is similar to that of the samples at high redshift
$z>1$ (cf. Fig.~\ref{fig:fig5}).

For the redshift outputs $z<0.5$, the relatively large number of dark
matter haloes allows us to split the catalogues into several
non-overlapping subsamples  having a number density $\nh\simeq
10^{-4}\hhhmmpc$. For these snapshots, we consider the mass bins
$10^{13}<M<1.6\times 10^{13}\mdh$, $1.6\times 10^{13} <M<3\times
10^{13}\mdh$ and $M>3\times 10^{13}\mdh$.  Results are shown in
Fig.~\ref{fig:fig6}. An increase in the ratio $\Delta b^{s}/\Delta
b^{t}$ as a function of wavenumber followed by a change of sign can
also be seen in these low biased samples in spite of the noisier data.
Notice that the linear halo bias is in the range $1\lsim b(M)\lsim
2$. In particular, the $z=0$ haloes with mass $10^{13}<M<1.6\times
10^{13}\mdh$ constitute an almost unbiased sample of the density
field, with $b(M)\approx 1.10$. At scales $k\lsim 0.02\hmmpc$ there is
some evidence that the non-Gaussian bias correction measured in the
low biased samples may be larger than the theoretical
expectation. Still, the bias shift is quite small for $b(M)=1.10$, in
agreement with the theoretical prediction that the effect  vanishes
for $b(M)=1$ assuming the Eulerian bias prescription $b(M)=1+b_{\rm
L}(M)$ (where $b_{\rm L}(M)$ is the Lagrangian bias) used in
eq.~(\ref{eq:dbias}).  Unfortunately, our simulations do not have
sufficient mass resolution to  resolve anti-biased haloes with
$b(M)<1$, for which theoretical predictions  based on the
peak-background split suggest the sign of the scale-dependent
contribution $\Delta b_\kappa$ is reversed.

We have not examined the behaviour of $\Delta b^{s}/\Delta b^{t}$ at
$k>0.1\hmmpc$ since the effect is already quite small there and
nonlinear bias due to galaxy evolution effects dominates. Most of  the
information on the  non-Gaussian bias comes from measurements at large
scale $k\lsim 0.03\hmmpc$ (see Slosar \etal 2008), where the
theoretical model and the numerical data agree reasonably well with
each other over the relevant range $2<b<3$.

\begin{figure}
\center 
\resizebox{0.5\textwidth}{!}{\includegraphics{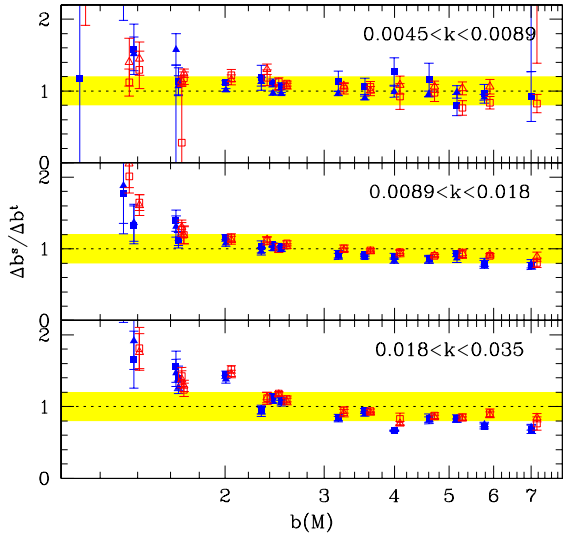}}
\caption{Ratio of simulations to theoretical predictions, 
eq.~(\ref{eq:dbias}),  as a function  of linear halo bias. The
scale-dependent correction is calculated using all  the wavemodes in
the wavenumber ranges quoted on the figure.  Squares and triangles
indicate the value of the ratio $\Delta b^{s}/\Delta b^{t}$ calculated
from $\phh$ and $\pmh$, respectively.  The measurements from the
non-Gaussian simulations with $\fnl=+100$  are marked as filled
symbols. The error bars are computed from a jackknife estimate.}
\label{fig:fig8}
\end{figure}

To reduce the scatter in the measurement of $\Delta b(k,\fnl)$, we can
increase the bin width $\Delta\log k$ so as to increase the number of
independent modes. In Fig.~\ref{fig:fig8}, the ratio  $\Delta
b^{s}/\Delta b^{t}$  is shown as a function of the linear halo bias
for three equally spaced logarithmic interval spanning the range
$0.0045<k<0.035\hmmpc$ (e.g. $\Delta\log k=0.3$).  The data points are
harvested from several outputs spanning the redshift range $0<z<2$
(i.e., the snapshot redshifts are $z=0$, 0.3, 0.5, 1, 1.4, 1.7 and 2).
The squares and triangles represent the deviation from the theoretical
prediction obtained by taking ratios of $\pmh$ and $\phh$,
respectively.  Filled symbols show results for the non-Gaussian
simulations with $\fnl=+100$. The error bars indicate our jackknife
error estimates on the average $\Delta b^{s}/\Delta b^{t}$.

The non-Gaussian bias shift of the low biased samples, $b<2$, appears
to deviate from the theory.  Equation~(\ref{eq:dbias}) may thus  need
correction when the linear bias gets lower than $\lsim 1.5-2$.  The
scale-dependent bias is quite large around $b \sim 1.3$, although for
$b \sim 1$ the effect does appear to vanish on the largest scales as
expected (see the upper left panel of Fig.~\ref{fig:fig7}).  For the
haloes with $b(M)\bsim 2$, the theory matches the non-Gaussian bias
correction for $k\lsim 0.01\hmmpc$.  One would need even larger
simulation boxes than used here to properly sample the largest
scales. For $k\bsim 0.01\hmmpc$, the effect is slightly suppressed
compared to the theoretical prediction, but this may plausibly arise
from  uncertainties in the magnitude of the theoretical
scale-independent shift  $\Delta b_{\rm I}$. Indeed, we have found
that a 20 per cent increase in $\Delta b_{\rm I}$  considerably
improves the agreement for $k\lsim 0.05\hmmpc$ and, at the same time,
is still consistent with the measured fractional  change in the
multiplicity function (see Fig.~\ref{fig:fig1}).  Finally, note that
that haloes with similar bias also have a comparable scale-dependent
bias  due to non-gaussianity regardless of redshift. Hence, there  is
no need to introduce a second parameter such as redshift for the
purpose of describing these results. 

The only previous work with simulations along these lines is that of
Dalal \etal (2008). These authors do not include the scale-independent
shift $\Delta b_{\rm I}$ nor the weaker correction $b(M)\beta_{\rm m}$
induced by the matter power spectrum. Hence, Fig.~8 of their paper
indeed shows $\Delta b^{s}/\Delta b_\kappa$. This ratio appears to
increase with wavenumber (even though their data points do not extend
beyond $0.03\hmmpc$), while the bottom panel of our
Fig.~\ref{fig:fig4} shows that $\Delta b^{s}/\Delta b_\kappa$ is
suppressed at high wavenumbers. Note, however, that their theoretical
scale-dependent correction  $\Delta b_\kappa(k,\fnl)$ does not include
the matter transfer function.  We found that, if the transfer function
were removed from eq.~(\ref{eq:bshift}), $\Delta b^{s}/\Delta
b_\kappa$ would be enhanced rather than suppressed as one goes to
higher wavenumber, in qualitative agreement with their findings.
Since the  non-Gaussianity is imprinted in the initial conditions
prior to the  evolution through  matter and radiation domination, the
transfer  function must be included in the analysis.

\subsection{Scaling with $\fnl$}

\begin{figure}
\center \resizebox{0.45\textwidth}{!}{\includegraphics{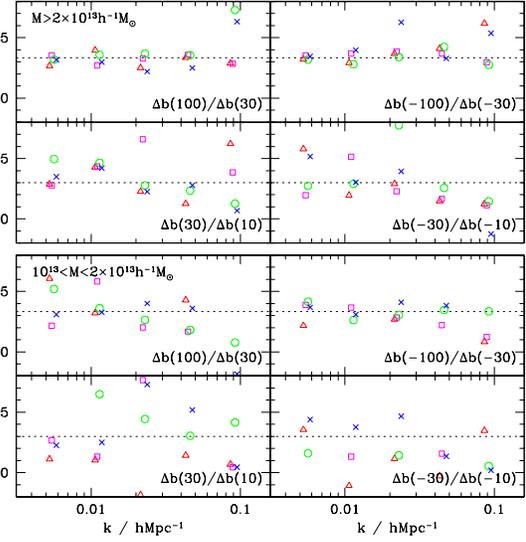}}
\caption{Sensitivity of the non-Gaussian shift to the strength of the
nonlinear parameter $\fnl$. The ratio  $\Delta b(k,\fnl^1)/\Delta
b(k,\fnl^2)$ is plotted as a function of wavenumber for various values
of $\fnl^1$ and $\fnl^2$ spanning the range [-100,+100]. Symbols show
results at $z=0.3$ (triangle) 0.5 (square), 1 (circle) and 2 (cross)
for two different halo mass cuts~:  $M>2\times 10^{13}\mdh$ (upper
panels) and $1<M<2\times 10^{13}\mdh$.  The horizontal lines indicate
the linear scaling $\fnl^1/\fnl^2$.}
\label{fig:fig9}
\end{figure}

The quadratic term $\fnl\phi^2$ also induces second and higher order
corrections to the effective bias shift $\Delta b(k,\fnl)$ which may
become important at high wavenumber. To test for these high order
terms, we explore in Fig.~\ref{fig:fig9} the scaling of the
non-Gaussian bias shift with the strength of the nonlinear parameter
$\fnl$.  Symbols show the ratio $\Delta b(k,\fnl^1)/\Delta
b(k,\fnl^2)$ (which we abridge $\Delta b(\fnl^1)/\Delta b(\fnl^2)$ for
shorthand convenience) as a function of wavenumber and redshift for
several values of $\fnl^1$ and $\fnl^2$ spanning the range
[-100,+100], as indicated in the figure.  Note that the data points
are obtained by averaging over two realisations only. The horizontal
line indicates  the value $\fnl^1/\fnl^2$ that should be reached if
the non-Gaussian  bias shift is linear in $\fnl$.

As we can see, there is less scatter in  $\Delta b(\pm 100)/\Delta
b(\pm 30)$ than in $\Delta b(\pm 30)/\Delta b(\pm 10)$ but, in both
cases, the results are broadly consistent with the linear expectation
$\fnl^1/\fnl^2$.  Furthermore, there is no significant dependence on
the wavenumber, redshift or the halo mass cut.
We conclude that the sensitivity of large scale structure bias should
extend to smaller values of $\fnl$ as expected. 

\subsection{Non-Gaussian bias in configuration space}

Thus far, we have investigated the impact of local non-Gaussianity on
two-point statistics in Fourier space. It is also instructive to
consider  the two-point correlation $\xi(r)$ in configuration space, 
which is related to the power spectrum $P(k)$ through
\begin{equation}
\xi(r)=\frac{1}{2\pi^2}\int_0^\infty\!\!\dd k k^2 P(k) j_0(kr)\;,
\end{equation}
where $j_0(x)$ is the zeroth spherical Bessel function. In practice,
since the simulation volume is a periodic cube, we compute the 
correlation from a discrete Fourier transform of the power spectrum.

\begin{figure}
\center \resizebox{0.45\textwidth}{!}{\includegraphics{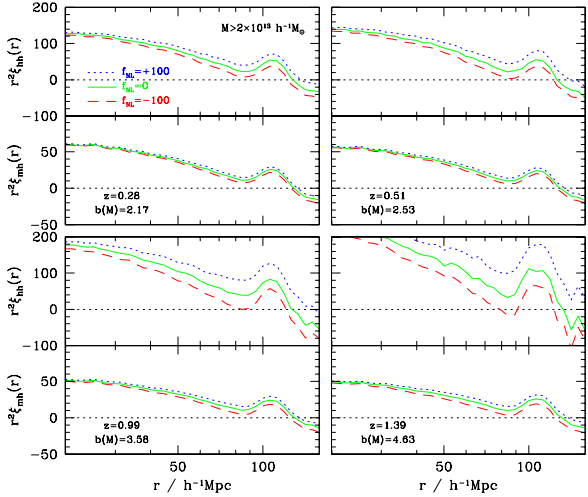}}
\caption{Effect of local non-Gaussianity on the auto and
cross-correlation function of haloes and dark matter, $\xi_{\rm
hh}(r)$ and $\xi_{\rm mh}(r)$. Results are shown as a function of
comoving  separation $r$ for the samples with $M>2\times 10^{13}\mdh$.
The respective values of output  redshift and linear halo bias are 
also quoted. }
\label{fig:fig10}
\end{figure}

In Fig.~\ref{fig:fig10}, the result of measuring the auto and
cross-correlation functions is shown at $0.3<z<1.5$ for the mass cut
$M>2\times 10^{13}\mdh$. The width of the simulation box is large
enough to sample wavemodes relevant to the baryon acoustic
oscillations (BAO).  The interesting feature of Fig.~\ref{fig:fig10}
is the correlation between the BAO and the broadband power, which
shows up differently in the correlation function than in the power
spectrum. Local non-Gaussianity adds broadband power and, therefore,
modulates the amplitude of the  BAO and the position of zero-crossing.

\section{Discussion and conclusions}
\label{sec:conclusion}

The scale dependence of clustering of biased tracers of the density
field has emerged  as a powerful method to constrain the amount of
primordial non-Gaussianity of the local type. In this paper, we have
measured the non-Gaussian bias correction $\Delta b(k,\fnl)$ in the
clustering of dark matter haloes extracted from a suite of large
N-body simulations.  In contrast to previous work, we  focus both on
the halo-halo and halo-matter power spectrum.  While we confirm the
basic effect reported in Dalal \etal (2008), we emphasize the
importance of including a scale-independent term $\Delta b_{\rm I}$
and, to a lesser extent, a contribution induced by the matter power
spectrum $b(M)\beta_{\rm m}$, to the scale-dependent shift $\Delta
b_\kappa$ when comparing the theoretical scaling to numerical
simulations. The inclusion of these two first order corrections
significantly improves the agreement at wavenumber  $k\lsim 0.1\hmmpc$.

The  original analysis in Dalal \etal (2008) only used cross-power
spectra  from simulations, while the data analysis in Slosar \etal
(2008) used  mostly auto-power analysis.  The two do not have to agree
with each other if the haloes and dark matter  do not trace each other
on large scales, i.e. if there is stochasticity.  While models with
Gaussian initial conditions predict there is little  stochasticity on
large scales (Seljak \& Warren 2004), this has not  been shown
explicitly for models with non-Gaussianity.  Hence, one of the main
motivations for this work was to extract the  non-Gaussianity effect
from the auto-correlations.  Measurements of the non-Gaussian bias
correction  obtained with the halo-halo or the halo-matter power
spectrum are in a good agreement with each other, indicating that
non-Gaussianity does not induce  stochasticity and the predicted
scaling applies equally  well for the auto- and cross-power spectrum.
The issue of stochasticity in non-gaussian models will be explored 
further in a future publication. 

For biased haloes ($b\bsim 1.5$), our results indicate that the
simulated non-Gaussian bias converges towards the theoretical
prediction  for $k\lsim 0.03\hmmpc$. At smaller scales, the effect 
depends on scale-independent bias. If it is ignored then the amplitude 
of the effect is suppressed relative to theory. If we include scale 
independent bias using analytic calculation this suppression is much 
smaller and in some cases goes in the opposite direction. Moreover, 
one could argue that scale-independent bias cannot be identified from 
the data alone, so one should fit for it and include it in the 
overall bias, as was done in 
Slosar \etal 2008. In this case the agreement between theory and 
simulations is improved further. Still, there is some evidence that 
for very biased haloes, $b>3$, the effect is suppressed relative to 
theory even on very large scales. 

For the halo samples with $b(M)\lsim 1.5$ there is some evidence that
the actual bias exceeds the theory on all scales. 
Therefore, the proposed eq.~(\ref{eq:dbias}) does not
appear to be universal,  so care must be exercised when applied to the
actual large scale structure data. It would be useful to verify
eq.~(\ref{eq:dbias}) on dark matter haloes which are anti-biased
($b(M)<1$) relative to the matter distribution, to see if the sign of
the effect is reversed. One, however, needs a very large volume and  a
very high mass resolution, which prevents us from verifying the
predictions in this regime with the current simulations.

On the observational side, Slosar \etal (2008) have already applied
the method to a sample of highly biased LRGs and QSOs, with mean bias
$b(M)\sim 1.8$ and 2.7, respectively. It is interesting to inspect how
those constraints change in light of our analysis. Our results suggest 
that for these values of halo bias theory and simulations are 
largely in agreement on relevant scales:
their constraints arise mostly from  the measurement of the
quasar power spectrum with $b \sim 2.7$ at the largest angular scales,
$k\lsim 0.005\hmmpc$ and from LRGs with $b \sim 1.8$ at $k\lsim
0.01\hmmpc$. As we see from Fig. \ref{fig:fig8}, theoretical
predictions  are in very good agreement with the simulations for these
values of bias and scales. Hence, we thus expect their limits remain
unchanged. 

Finally, we note that we have not considered other effects that may
also  modify the predictions, such as redshift space distortions and
merger bias. The latter can significantly  weaken the
predicted scale-dependent bias (Slosar \etal 2008). We plan  to
investigate these effects with simulations in the future.

\section*{Acknowledgements}

We are indebted to Volker Springel and Alexander Knebe for making
their code, resp. {\small GADGET2} and {\small AMIGA}, available.  We
thank Pat McDonald and Robert Smith for useful discussions and Chris
Hirata for pointing out to us the importance of scale-independent
bias.  The simulations used in this paper were run on the zBOX3
supercomputer at the University of Z\"urich.  We acknowledge support
from the Swiss National Foundation (Contract No. 200021-116696/1).

\appendix

\section{Perturbation theory with local non-Gaussianity}
\label{app:PT}

\subsection{Skewness parameter}

In $\fnl$ non-Gaussianity, the Fourier mode of the curvature perturbation
(after matter-radiation equality) is given by
\begin{equation}
\Phi(\vk)=\phi(\vk)+\fnl\int\!\!\frac{\dd^3q}{(2\pi)^2}\phi(\vq)
\phi(\vk-\vq)\;,
\end{equation}
where $\phi$ is the unperturbed Gaussian field with power spectrum 
$P_\phi(k)\propto k^{n_s-4}$. The primordial bispectrum of curvature 
perturbations is
\begin{equation}
B_\Phi(k_1,k_2,k_3)=2\fnl\left[P_\phi(k_1)P_\phi(k_2)+2~\mbox{perms}~
\right]\;.
\end{equation}
Hence, the three-point correlation of the Fourier modes of the smoothed
matter density field, $\delta_M(\vk)=\alpha(M,k)\Phi(\vk)$, reads
\begin{eqnarray}
\la\delta_M(\vk_1)\delta_M(\vk_2)\delta_M(\vk_3)\ra \!\!\!&=&\!\!\!
(2\pi)^3\,\alpha_1\alpha_2\alpha_3\,B_\Phi(k_1,k_2,k_3) \nonumber \\ 
&& \times \delta_{\rm D}(\vk_1+\vk_2+\vk_3) \;,
\end{eqnarray}
where $\alpha_i=\alpha(M,k_i)$ for shorthand convenience. Here, 
$\delta_{\rm D}$ is the Dirac delta and the transfer function $\alpha(M,k)$ 
is given by eq.~(\ref{eq:alpha}). Note that we have omitted the explicit 
redshift dependence of $\delta_M$ and $\alpha$ for brevity. The (connected) 
three-point function of $\delta_M$ in configuration space is the Fourier 
transform of $\la\delta_M(\vk_1)\delta_M(\vk_2)\delta_M(\vk_3)\ra$. In 
particular, the third-moment of the smoothed density field is 
\begin{eqnarray}
\la\delta_M^3\ra \!\!\!&=&\!\!\!\int\!\!\frac{\dd^3 k_1}{(2\pi)^3}
\int\!\!\frac{\dd^3 k_2}{(2\pi)^3}\int\!\!\frac{\dd^3 k_3}{(2\pi)^3}
\la\delta_M(\vk_1)\delta_M(\vk_2)\delta_M(\vk_3)\ra \nonumber \\
&=& \!\!\! 2\fnl\int\!\!\frac{\dd^3 k_1}{(2\pi)^3}\int\!\!\frac{\dd^3 k_2}
{(2\pi)^3}\,\alpha_1\alpha_2\alpha_3\,P_\phi(k_1)P_\phi(k_2) \nonumber \\
&& \!\!\! \times \left[1+\frac{P_\phi(k_3)}{P_\phi(k_1)}
+\frac{P_\phi(k_3)}{P_\phi(k_2)}\right]\;.
\end{eqnarray}
We have used the momentum conservation implied by the Dirac delta, i.e. 
$\vk_3=-\vk_1-\vk_2$, to obtain the second line. Eq.~(\ref{eq:skew})
follows after taking advantage of the invariance under the exchange
of $\vk_1$ with $\vk_2$ and integrating out some of the angular variables.

\subsection{Matter power spectrum}

Following Taruya \etal (2008), we estimate the non-Gaussian correction 
to the matter power spectrum in the weakly nonlinear range,
$k\lsim 0.1\hmmpc$, using perturbation theory. At the first order, the
matter power spectrum can be expressed as
\begin{eqnarray}
\pmm(k,\fnl) \!\!\! &=& \!\!\! 
D^2(z) \pl(k) + \left[P^{(22)}(k,z)+P^{(13)}(k,z)\right]
\nonumber \\ 
&& + P^{(12)}(k,z;\fnl)\;.
\end{eqnarray}
Here, $D(z)$ is the growth factor, $P_{\rm L}(k)$ is the linear power
spectrum of the density field,  
\begin{eqnarray}
P^{(22)}(k,z) \!\!\! &=& \!\!\! D^4(z)\, \frac{k^3}{98(2\pi)^2}
\int_0^\infty\!\!\dd x\, \pl(kx) \\
&& \times\int_{-1}^{+1}\!\!\dd\mu\,\pl(k\sqrt{1+x^2-2\mu x}) \nonumber \\
&& \times\left(\frac{3x+7\mu-10\mu^2 x}{1+x^2-2\mu x}\right)^2  \nonumber \\
P^{(13)}(k,z) \!\!\! &=& \!\!\! D^4(z)\, \frac{k^3\pl(k)}{252(2\pi)^2}
\int_0^\infty\!\!\dd x\,\pl(kx) \\ 
&& \times \left[\frac{12}{x^2}-158+100 x^2-42 x^4 \right. \nonumber \\ 
&& \left. + \frac{3}{x^3}\left(x^2-1\right)^2\left(7 x^2+2\right)
\ln\left|\frac{1+x}{1-x}\right|\right] \nonumber
\end{eqnarray}
are the standard one-loop contributions in the case of Gaussian initial
conditions (e.g. Goroff \etal 1986; Makino \etal 1992; Jain \& Bertschinger 
1994) and
\begin{eqnarray}
P^{(12)}(k,z;\fnl) \!\!\! &=& \!\!\! \fnl\frac{2 k^3}{7(2\pi)^2}\,\alpha(0,k)
\int_0^\infty\!\!\dd x\,x\, \alpha(0,kx) \nonumber \\
&& \times\int_{-1}^{+1}\!\!\dd\mu\,\left(\frac{3x+7\mu-10\mu^2 x}
{1+x^2-2\mu x}\right)\alpha(0,q) \nonumber \\
&& \times \left[P_\phi(k) P_\phi(kx)+\mbox{ 2 perms }\right] \;,
\end{eqnarray}
where $q^2=k^2(1+x^2-2\mu x)$, is the leading-order correction due to
local  non-Gaussianity which arises from the non-zero primordial
bispectrum of  curvature perturbations. $\alpha(0,k)$ is the function
eq.(\ref{eq:alpha}) with filtering kernel $W(0,k)\equiv 1$. The
particular redshift dependence of these power spectra follows from
the assumption of growing-mode initial conditions. The relative
contribution $\beta_{\rm m}(k,\fnl)=\Delta\pmm(k,\fnl)/\pmm(k,0)$ of
local non-Gaussianity thus is
\begin{equation}
\beta_{\rm m}(k,\fnl)=\frac{P^{(12)}(k,z;\fnl)}
{D^2(z) \pl(k)+P^{(22)}(k,z)+P^{(13)}(k,z)}
\end{equation}
at leading order.  Notice that this ratio scales as $\propto D(z)$, so
the effect of  local non-Gaussianity on the matter power spectrum is
largest at  low redshift.

\label{lastpage}

\end{document}